\documentclass[conference]{IEEEtran}
\IEEEoverridecommandlockouts
\usepackage{cite}
\usepackage{amsmath,amssymb,amsfonts}
\usepackage{algorithmic}
\usepackage{subfig}
\usepackage{graphicx}
\usepackage{textcomp}
\usepackage{xcolor}
\usepackage[a4paper, total={184mm,239mm}]{geometry}
\def\BibTeX{{\rm B\kern-.05em{\sc i\kern-.025em b}\kern-.08em
    T\kern-.1667em\lower.7ex\hbox{E}\kern-.125emX}}

\usepackage{mathptmx} 
\usepackage{color}
\usepackage[normalem]{ulem}
\usepackage{tikz}
\definecolor{yellow}{RGB}{255,217,101}      
\definecolor{blue}{RGB}{102,153,255}      
\definecolor{green}{RGB}{146,208,80}      
\newcommand*\circled[1]{\tikz[baseline=(char.base)]{
            \node[text = white, inner color = black, outer color = black, shape=circle,draw,inner sep=0.3pt] (char) {#1};}}
            
\newcommand*\circledy[1]{\tikz[baseline=(char.base)]{
            \node[text = black, inner color = yellow, outer color = yellow, shape=circle,draw,inner sep=0.3pt] (char) {#1};}}

\usepackage{makecell}
\usepackage{multirow}
\usepackage{marvosym}
\usepackage{fancyhdr}
\usepackage{graphicx}
\usepackage{array}
\newcommand{\preserveBackslash}[1]{\let\temp=\\#1\let\\=\temp}
\newcolumntype{C}[1]{>{\preserveBackslash\centering}p{#1}}

\usepackage[colorlinks,
            linkcolor=blue,       
            anchorcolor=blue,  
            citecolor=blue,        
            ]{hyperref}
\begin{document}

\title{Hybrid Photonic-digital Accelerator for Attention Mechanism\\
}

\author{\IEEEauthorblockN{Huize~Li, Dan Chen\textsuperscript{\Letter}, Tulika Mitra}
\IEEEauthorblockA{School of Computing, National University of Singapore, 119077, Singapore\\
huizeli@nus.edu.sg, danchen@nus.edu.sg (Corresponding author), tulika@comp.nus.edu.sg}}

\maketitle

\begin{abstract}

The wide adoption and substantial computational resource requirements of attention-based Transformers have spurred the demand for efficient hardware accelerators. Unlike digital-based accelerators, there is growing interest in exploring photonics due to its high energy efficiency and ultra-fast processing speeds. However, the significant signal conversion overhead limits the performance of photonic-based accelerators. In this work, we propose HyAtten, a photonic-based attention accelerator with minimize signal conversion overhead. HyAtten incorporates a signal comparator to classify signals into two categories based on whether they can be processed by low-resolution converters. HyAtten integrates low-resolution converters to process all low-resolution signals, thereby boosting the parallelism of photonic computing. For signals requiring high-resolution conversion, HyAtten uses digital circuits instead of signal converters to reduce area and latency overhead. Compared to state-of-the-art photonic-based Transformer accelerator, HyAtten achieves 9.8$\times$ performance/area and 2.2$\times$ energy-efficiency/area improvement.

\end{abstract}


\section{Introduction}

Transformer-based neural networks have achieved remarkable success in various domains, such as {\em natural language processing} (NLP)~\cite{Bai24} and {\em computer vision} (CV)~\cite{You23}. The core operation of Transformers is the self-attention mechanism, which calculates pairwise correlations between input tokens to enhance inference accuracy. Despite their superior accuracy, the quadratic complexity of self-attention requires substantial computational resources, posing a significant challenge for deploying Transformers, especially in resource-constrained systems. Consequently, there is a pressing need to develop domain-specific hardware accelerators to enable the efficient deployment of Transformers in real-world applications.

Several digital hardware accelerators have been proposed to improve the inference performance of Transformers by reducing redundant memory access and enhancing computational parallelism~\cite{You23, Bai24, Li24, sadimm24}. While these digital accelerators effectively reduce inference latency, traditional electrical computing platforms face significant limitations as transistor-based chips approach the boundaries of Moore's Law. This results in increased power dissipation, particularly in computation-intensive self-attention processes. As Transformer models continue to grow in size, the high latency and energy consumption faced by digital accelerators will only become more pronounced. In contrast, integrated photonic accelerators present a promising alternative for accelerating deep neural networks, offering ultra-high speeds, extensive parallelism, and low energy consumption.

Various optical systems have been explored to accelerate {\em convolutional neural networks} (CNNs)\cite{Feldmann21} and Transformers\cite{Zhu24}. However, existing photonic accelerators are highly dependent on high-resolution signal converters to preserve the accuracy of neural network inference, as shown in Figure~\ref{motiv} (a). These high-resolution converters, however, have become a substantial bottleneck, constraining the overall performance of photonic accelerators. For instance, in the state-of-the-art photonic-based Transformer accelerator, Lightening-Transformer~\cite{Zhu24}, signal conversion units, {\em analog-to-digital converters} (ADCs), {\em digital-to-analog converters} (DACs), and optical/electrical converters, consume more than 50\% of the chip's area. To mitigate the area overhead, Lightening-Transformer shares one ADC among multiple photonic arrays. However, one 32$\times$32 photonic array will generate 1024 signals in one cycle, which can not be efficiently processed by only one ADC. Hence, there is a pressing need to minimize the delay, energy, and area overhead of signal conversion units without sacrificing model accuracy.

Given this context, we introduce HyAtten, a novel photonic-based attention mechanism accelerator designed to minimize signal conversion overhead while maintaining inference accuracy. Our experimental results reveal that over 85\% of the analog signals in existing photonic-based Transformer accelerators can be effectively processed using low-resolution converters, such as 4-bit ADCs. Leveraging this insight, HyAtten utilizes low-resolution converters within its photonic circuits to handle these signals, substantially reducing both the latency and area overhead of the photonic components. For the remaining high-resolution signals (exceed the full-scale measurement range of low-resolution converters) that cannot be processed by the low-resolution converters, HyAtten integrates digital circuits to manage these part of attention computations. Importantly, the digital circuits only need to process a small fraction of the data (less than 15\%), significantly lowering the overall demand for digital computational resources. The key contributions of HyAtten are as follows:

\begin{itemize}
    \item We perform extensive experiments on a state-of-the-art photonic Transformer accelerator and find that over 85\% of analog signals can be efficiently processed using low-resolution signal converters.
    \item Based on this insight, we propose HyAtten, an innovative photonic-based attention mechanism accelerator that employs low-resolution converters to handle these signals.
    \item To further optimize performance, HyAtten incorporates digital circuits to process the remaining high-resolution signals, minimizing signal conversion overhead.
    \item Our evaluations demonstrate that HyAtten outperforms existing photonic Transformer accelerators, achieving a 9.8$\times$ improvement in performance-per-area and a 2.2$\times$ increase in energy-efficiency-per-area.
\end{itemize}

\section{Background and Motivation}
\subsection{Transformer and self-attention}

Transformer-based neural networks are generally composed of multiple identical blocks, referred to as encoder and decoder blocks. Both types of blocks contain a {\em multi-head self-attention} (MHA) module, a {\em feed-forward network} (FFN), shortcut connections, and {\em layer normalization} (LN). Additionally, the decoder block incorporates cross-attention and masked self-attention modules. For illustration, the structure of a basic encoder block is defined as follows:

\begin{equation}
\begin{cases}
{\bf X}_{l+1}^{'} = {MHA}({LN}({\bf X}_l)) + {\bf X}_l; \\
{\bf X}_{l+1} = {FFN}({LN}({\bf X}_{l+1}^{'})) + {\bf X}_{l+1}^{'}, \\
\end{cases}
\label{exp1}
\end{equation}
where ${\bf X}_{l}$ is the input sequences of $l$-th layer.

{\em Multi-head Self-Attention} (MHA) mechanism consists of $H$ distinct self-attention heads. Within each head, the input vector is linearly projected into three separate vectors: the query ({\bf Q}), key ({\bf K}), and value ({\bf V}) vectors. The attention function is then computed between these input vectors as follows:

\begin{equation}
Attention({\bf Q}, {\bf K}, {\bf V}) = softmax({\bf QK}^{\mathsf{T}} / \sqrt{d_{k}}){\bf V},
\label{exp2}
\end{equation}
where $d_k$ is {\bf Q} and {\bf K}'s dimension. An intermediate score matrix {\bf S} is obtained with $S = Q\times K^\mathsf{T}$. As the input sequence grows, {\bf Q}, {\bf K}, and {\bf V} will grow linearly while the attention computation of matrix {\bf S} will grow quadratically. The FFN module usually contains two linear layers with an activation function in between.

\begin{figure}[t]
\centering
\includegraphics[width=8.0cm]{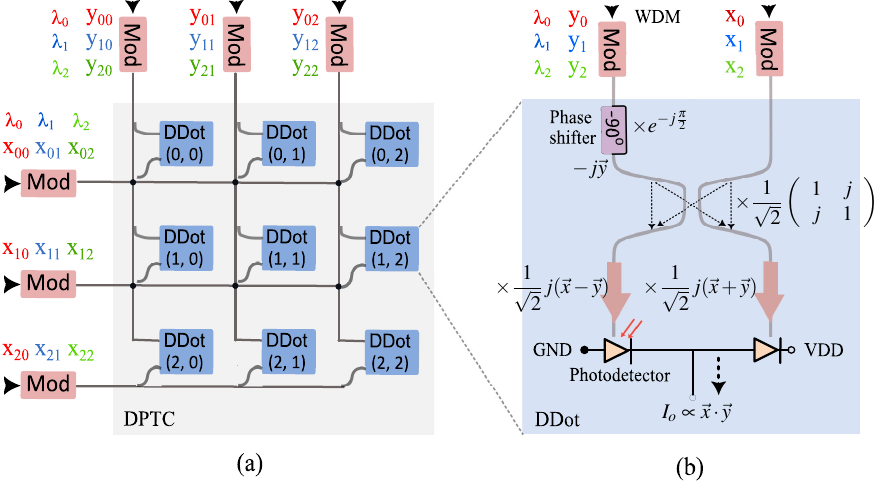}

\caption{(a) DPTC array proposed in~\cite{Zhu24}, and (b) DDot unit proposed in~\cite{Zhu24}}

\label{figure:basic}
\end{figure}

\subsection{Optical computing basics}
Recently, researchers propose a {\em dynamically-operated dot-product} (DDot) unit to perform optical dot-products between two vectors $\Vec{x}$ and $\Vec{y}$~\cite{Zhu24}. As Figure~\ref{figure:basic} (b) shows, the DDot units are designed based on coherent interference. First, the {\em wavelength-division multiplexing} (WDM) technique encodes each input pairs ($x_i$, $y_i$) in the same wavelength $\lambda_i$. The WDM light signals are then sent through the two arms of $50:50$ {\em directional coupler} (DC) with a $-90^{\circ}$ {\em phase shifter} (PS). Consequently, the two output signals become orthogonal in the complex plane. This setup allows each input pair ($x_i$, $y_i$) with the same wavelength $\lambda_i$ to interfere in parallel, while different wavelengths do not interfere. The photo-diode at the end of each output port of DC converts the incident WDM signals into photocurrent. The generated photocurrent is proportional to the accumulated intensities of the WDM signals, representing the square of the optical magnitudes and producing the final output current as $I_o \propto \Vec{x}\cdot\Vec{y}$.

To enable {\em general matrix multiplication} (GEMM) operations with optical dot-product engines, a {\em dynamically-operated photonic tensor core} (DPTC) has been proposed~\cite{Zhu24}. Researchers designed a compact crossbar array of DDot units to maximize operand sharing within the core, significantly reducing operand modulation costs, as illustrated in Figure~\ref{figure:basic} (a). This architecture supports efficient sharing of photonic waveguide buses among DDot units, enabling ultra-parallel GEMM operations. A $N_v \times N_h$ DPTC comprises $N_v \times N_h$ DDot units, where $N_v$ and $N_h$ denote the numbers of input waveguides in the vertical and horizontal directions, respectively.

\begin{figure}[t]
\centering
\subfloat[]{
\begin{minipage}[t]{0.489\linewidth}
\centering
\includegraphics[width=1.55in]{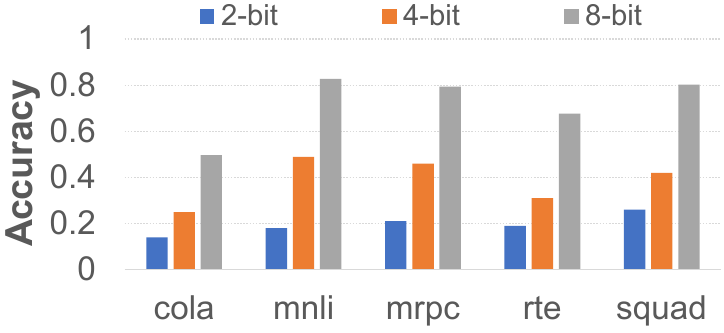}
\end{minipage}
}%
\subfloat[]{
\begin{minipage}[t]{0.49\linewidth}
\centering
\includegraphics[width=1.5in]{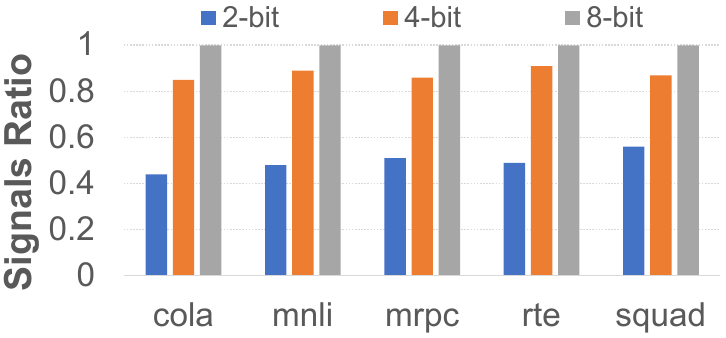}
\end{minipage}
}%
\centering

\caption{(a) The model accuracy when employing different ADC resolutions, and (b) the proportion of signals that remain within the resolution limits of various ADCs}

\label{motiv}

\end{figure}

\subsection{Motivations}
\label{sec:motiv}
{\em Problem: Signal conversion costs remain the primary bottleneck for emerging photonic systems.} In the state-of-the-art photonic Transformer accelerator, Lightening-Transformer~\cite{Zhu24}, signal conversion units, such as ADCs, DACs, and optical/electrical converters, consume over 50\% of the chip area. To reduce this overhead, Lightening-Transformer allocates a single ADC to a 32$\times$32 DPTC array. However, a 32$\times$32 DPTC array generates 1024 analog signals, which cannot be efficiently processed by just one ADC. As a result, photonic devices experience significant delays, idling while awaiting signal conversion.

{\em Observation\#1: Utilizing low-resolution signal converters can significantly reduce latency and area overhead, but it may also lead to substantial model accuracy loss.} According to the latest ADC performance survey~\cite{Murmann2023}, published in May 2023, ADC area increases exponentially with resolution. For example, a 5-bit ADC requires twice the area of a 4-bit ADC using the same technology. While a common approach to reduce area overhead is to employ low-resolution ADCs, we evaluated the impact of varying ADC resolutions (2-bit, 4-bit, and 8-bit) on the Lightening-Transformer~\cite{Zhu24}. Our evaluation utilized five datasets, CoLA, MNLI, MRPC, RTE, and SQuAD, from GLUE~\cite{Wang18glue}, running on a BERT-based model. For high-resolution output signals that exceed the ADC's full-scale measurement range, the conversion output was capped at the ADC's maximum value. The accuracy results (ratio between the number of correctly predicted samples by the total number of samples), shown in Figure~\ref{motiv} (a), reveal that using 4-bit ADCs leads to a more than 20\% decrease in model accuracy compared to the original 8-bit ADCs. These experiments demonstrate that high-resolution signals are critical to preserving model accuracy and must be appropriately handled.

{\em Observation\#2: Only a small fraction of analog signals require high-resolution ADCs.} Analogous to the barrel principle, ADC resolution must account for the ``short end of the barrel", i.e., the high-resolution signals~\cite{Zhu24}. If we can effectively address these high-resolution signals, the ADC resolution can be lowered for the remaining signals. Figure~\ref{motiv} (b) illustrates the proportion of signals exceeding the resolution of various ADCs from the above experiments. While 8-bit ADCs process all signals without exceeding their full-scale measurement range, 4-bit ADCs successfully process over 85\% of signals, leaving only 15\% unprocessed. This indicates that the ``barrel short" for 4-bit ADCs accounts for 15\% of the total signals.

{\bf Our goal:} Building on Observation\#2, analog signals in photonic-based Transformer accelerators can be categorized into two groups: low-resolution signals ($\leq$4-bit) and high-resolution signals ($>$4-bit). Low-resolution signals can be efficiently processed using 4-bit ADCs with lower latency and area overhead. As noted in Observation\#1, however, high-resolution signals require careful handling to avoid accuracy loss. Rather than introducing high-resolution ADCs, we employ digital circuits to process these signals. Since high-resolution signals constitute less than 15\% of the total, the computational overhead imposed on the digital circuits remains minimal.

\begin{figure}[t]
\centering
\includegraphics[width=9.0cm]{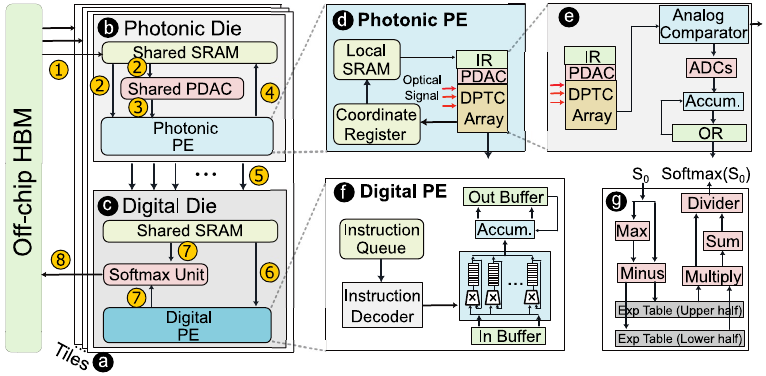}

\caption{Architecture and dataflow of HyAtten}

\label{figure:archi}
\end{figure}
\section{HyAtten}

\subsection{Architecture}
As depicted in Figure~\ref{figure:archi}~(\circled{a}), HyAtten consists of multiple Tiles, each of which includes a photonic die~(\circled{b}) and a digital die~(\circled{c}). The photonic die performs highly parallel photonic computing, while the digital die carries out digital computations without the need for signal conversion. Together, the photonic and digital dies collaborate to execute various components of the attention mechanism, such as GEMM and softmax operations. We assume that all data related to the attention mechanism, including matrices $Q$, $K$, and $V$, are stored in off-chip {\em High-bandwidth Memory} (HBM). The data will be transferred between the HBM and HyAtten Tiles.

{\bf Details of Photonic Die.} The photonic die includes a shared {\em static random access memory} (SRAM), a shared {\em photonic digital-to-analog converter} (PDAC) with modulation units, and a photonic {\em processing element} (PE)~(\circled{d}). Both the shared SRAM and PDAC are accessible by all Tiles. Input matrices $Q$ and $K$ from Equation~(\ref{exp2}) are transferred from the HBM to the shared SRAM. To reduce computational overhead, we apply low-bit quantization (4-bit) to these input matrices, as demonstrated in~\cite{Zhu24}, where low-bit quantization significantly decreases computational demands with minimal accuracy loss. To ensure the input matrices fit within the shared SRAM, HyAtten is designed to support batch-based processing, where matrices are partitioned into smaller batches (either row or column vectors) that are processed sequentially~\cite{Zhu24}. The shared PDAC receives one batch (e.g., one column vector) of the input matrix $Q$, converts it to photonic signals, and broadcasts these signals to all Tiles' photonic PEs for processing.

The photonic PE handles the core computations in the attention mechanism, specifically the GEMM operations for $Q\times K^\mathsf{T}$ and $S\times V$. As shown in Figure~\ref{figure:archi}~(\circled{d}), the photonic PE is equipped with local SRAM, a coordinate register, and a DPTC unit. The local SRAM receives data (e.g., matrix $K$) from the shared SRAM and transfers it to the DPTC units for processing. Unlike the shared SRAM, data in the local SRAM is routed to a local PDAC for signal conversion. The DPTC unit processes two sets of photonic inputs: one from the shared PDAC and another from the local PDAC, as illustrated in Figure~\ref{figure:basic} (a). The coordinate register stores the coordinates of some matrix elements, which are used to facilitate data loading from both the shared and local SRAM.

Figure~\ref{figure:archi}~(\circled{e}) illustrates the detailed architecture of a DPTC unit. Unlike prior photonic accelerators that configure one high-resolution ADC for each DPTC array~\cite{Zhu24}, we configure each 64$\times$64 DPTC array with 32 low-resolution ADCs to mitigate signal conversion latency. The DPTC unit operates as follows: two input photonic signals are received from the shared PDAC and local PDAC. These signals are processed through the DPTC array, resulting in photonic currents corresponding to the GEMM operations. Given the use of low-resolution ADCs, some photonic currents may exceed the ADCs' resolution. To handle this, an analog comparator is integrated to detect over-resolution signals and log their coordinates in the coordinate register. Under-resolution signals are processed by the ADCs to obtain their digital results. The memory controller then loads the over-resolution data based on the coordinates in the register and sends these data to the digital die for further processing.

{\bf Details of Digital Die.} The digital die consists of a shared SRAM, a softmax unit, and a digital PE. The shared SRAM receives two types of data from the photonic die: the GEMM output results of low-resolution signals and the input digital values of high-resolution signals. It also buffers the computed results before transferring them back to the HBM. The softmax unit handles the softmax operation for the attention score matrix $S$, which is collaboratively generated by the photonic and digital dies. The digital PE is responsible for completing GEMM operations that cannot be processed by the photonic die. The area and latency overhead of the digital die is minimal, as it processes only a small subset of signals.

Figure~\ref{figure:archi}~(\circled{f}) illustrates the architecture of the digital PE, which includes an instruction queue, an instruction decoder, an input buffer, a {\em multiplication-accumulation unit} (MAU), and an output buffer. The instruction queue and decoder work together to manage and schedule the digital PE's execution. Data is received from the shared SRAM by the input buffer, which then forwards it to the MAU. The MAU carries out vector-vector multiplications and accumulations to complete the GEMM operations. The results are subsequently stored in the output buffer for further processing or storage.

Figure~\ref{figure:archi}~(\circled{g}) presents the architecture of the softmax unit. Following the approach in\cite{cpsaa24}, we implement the exponent function using a lookup table. To minimize the size of the lookup table, we leverage the property that an exponentiation can be decomposed into the product of two smaller exponentiations. Thus, we employ two smaller lookup tables, an upper half and a lower half, and a multiplier to achieve the desired result. After computing the exponent of the dot-product, the value is accumulated and later used as the denominator in the softmax computation.

\begin{figure}[t]
\centering
\includegraphics[width=9.0cm]{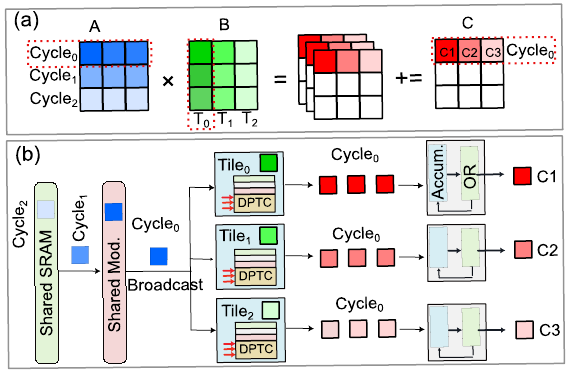}

\caption{GEMM operations on multiple photonic Tiles}

\label{figure:tile}
\end{figure}

\subsection{Dataflow}

{\bf Details of Transmission.} Figure~\ref{figure:archi} shows the data transmission process within HyAtten, using the $Q \times K^\mathsf{T}$ operation as an example, with a similar procedure followed for $S \times V$. In \circledy{1}, the input matrices $Q$ and $K$ are loaded from the off-chip HBM into the shared SRAM on HyAtten's photonic die. In \circledy{2}, matrix $K$ is transferred to the local SRAM of the photonic PE, while matrix $Q$ is sent to the shared PDAC for conversion into photonic signals. In \circledy{3}, the photonic signals corresponding to matrix $Q$ are broadcast to all Tiles' photonic PEs. Simultaneously, matrix $K$ is converted into photonic signals, and both matrices undergo GEMM operations to produce the result matrix $S$. In \circledy{4}, the result matrix $S$, along with the digital values of the high-resolution signals, are stored in the shared SRAM. The digital values of the high-resolution signals are transferred to the digital die in \circledy{5}. In \circledy{6}, the digital PE then processes these high-resolution signals, performing the remaining GEMM operations. In \circledy{7}, the matrix $S$ is sent to the softmax unit. The results will be stored back to the HBM in \circledy{8}.

{\bf Matrices partition and data mapping.} To address the size limitations of the DPTC array, the input matrices must be divided into smaller sub-matrices that align with the dimensions of the array. We adopt a Tile-based matrix partitioning and data mapping strategy, as illustrated in Figure~\ref{figure:tile} (a). The input matrices, $A$ and $B$, are partitioned into shards, with each shard matching the size of the DPTC array. Matrix $A$ is stored in the shared SRAM and accessed sequentially in a column-wise manner, while matrix $B$ is stored in the local SRAM of the photonic PE, with different shards distributed across different Tiles. For instance, as shown in Figure~\ref{figure:tile} (a), the first sub-matrix of matrix $B$ is stored in Tile$_0$. Given the limited capacity of the local SRAM, we process subsets of input matrices at a time. For example, if the local SRAM can accommodate two sub-matrices, GEMM operations will be executed in batches of two sub-matrices at a time.

\begin{table}[t]
\centering
\tabcolsep=0.12cm
    \caption{Hardware Configurations of HyAtten }

    \label{tab:breakdown}
    \begin{tabular}{|C{1.7cm}||C{1.5cm}|c|c|c|}  
    \cline{1-5}
    
       {\bf Component} & {\bf Area (mm$^2$) } & {\bf Power (mW)} & {\bf Params.} & {\bf Spec.}\\
       \hline \hline

       \multicolumn{5}{|c|}{Components Shared by all Tiles} \\ \hline
       \multirow{2}*{Shared PDAC} & \multirow{2}*{0.0016} & \multirow{2}*{8} & {Resolution} & {4 Bits} \\
       \cline{4-5}
       {} & {} & {} & {Numbers} & {1} \\
       \hline
       {Shared SRAM} & {3.68} & {1.23K} & {Capacity} & {2MB} \\
       \hline
       
       \multicolumn{5}{|c|}{Photonic Die (PD) Properties} \\ \hline
       \multirow{2}*{PDAC~\cite{Sridarshini20}} & \multirow{2}*{0.0748} & \multirow{2}*{520} & {Resolution} & {4 Bits} \\
       \cline{4-5}
       {} & {} & {} & {Numbers} & {64} \\
       \hline
       \multirow{2}*{ADC~\cite{Liu22}} & \multirow{2}*{0.0057} & \multirow{2}*{29.6} & {Resolution} & {4 Bits} \\
       \cline{4-5}
       {} & {} & {} & {Numbers} & {32} \\
       \hline
       \multirow{2}*{DPTC Array} & \multirow{2}*{0.246} & \multirow{2}*{624} & {Size} & {64 $\times$ 64} \\
       \cline{4-5}
       {} & {} & {} & {Numbers} & {1} \\
       \hline
       {SRAMs~\cite{Shafaei14}} & {0.06} & {19} & {Capacity} & {32KB} \\
       \hline
       {Registers} & {0.015} & {5.23} & {Capacity} & {8KB} \\
       \hline
       {Accumulator} & {0.0014} & {0.039} & {Numbers} & {32} \\
       \hline
       {Ana. Comp.} & {0.00031} & {0.019} & {Numbers} & {32} \\
       \hline
       {{\bf PD} Total} & {0.405} & {1.2K} & {Numbers} & {1} \\
       \hline\hline
       
       \multicolumn{5}{|c|}{Digital Die (DD) Properties} \\ \hline
       {MAU~\cite{Corti20}} & {0.014} & {8.2} & {Numbers} & {1} \\
       \hline
       {Registers} & {0.002} & {0.63} & {Capacity} & {1KB} \\
       \hline
       \multirow{2}*{Softmax~\cite{cpsaa24}} & \multirow{2}*{0.0072} & \multirow{2}*{1.134} & {LUT Size} & {512B} \\
       \cline{4-5}
       {} & {} & {} & {Numbers} & {1} \\
       \hline
       {{\bf DD} Total} & {0.023} & {9.96} & {Numbers} & {1} \\
       \hline\hline
       
        \multicolumn{5}{|c|}{HyAtten properties (32 Tiles in total)} \\ \hline
        {{\bf HyAtten}} & {17.38} & {39.9W} & {Numbers} & {1} \\
        \hline
\end{tabular}

\end{table}

{\bf Details of Calculation.} Figure~\ref{figure:tile} (b) illustrates the Tile-based GEMM operations. We assume that matrix $A$ has been partitioned and stored in the shared SRAM, while matrix $B$ has also been partitioned and stored in the local SRAM across different Tiles. During cycle$_0$, the first sub-matrix of matrix $A$ is sent to the shared PDAC, where it is converted into photonic signals and broadcast to all Tiles' photonic PEs. Simultaneously, the corresponding sub-matrix of matrix $B$ is converted to photonic signals by the local PDAC. The resulting photonic currents produced by the DPTC arrays are processed by the ADCs to generate the outputs of the sub-matrix multiplication. In the following two cycles, cycle$_1$ and cycle$_2$, the second and third sub-matrices of matrix $A$ are sequentially sent to the shared PDAC, where they are converted into photonic signals and broadcast to all Tiles. These signals multiply with the same sub-matrix of matrix $B$ as in cycle$_0$. The results from cycle$_0$ are accumulated to generate the first row vector of the output matrix, and this process continues for subsequent sub-matrix multiplications, following the same procedure.


\begin{figure}[t]
\centering
\subfloat[]{
\begin{minipage}[t]{0.495\linewidth}
\centering
\includegraphics[width=1.7in]{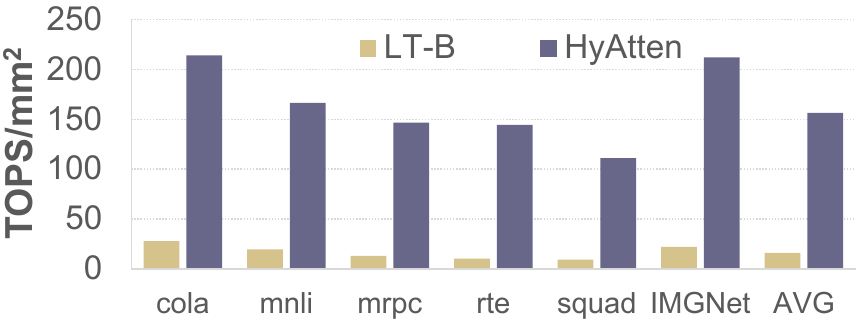}
\end{minipage}
}%
\subfloat[]{
\begin{minipage}[t]{0.495\linewidth}
\centering
\includegraphics[width=1.7in]{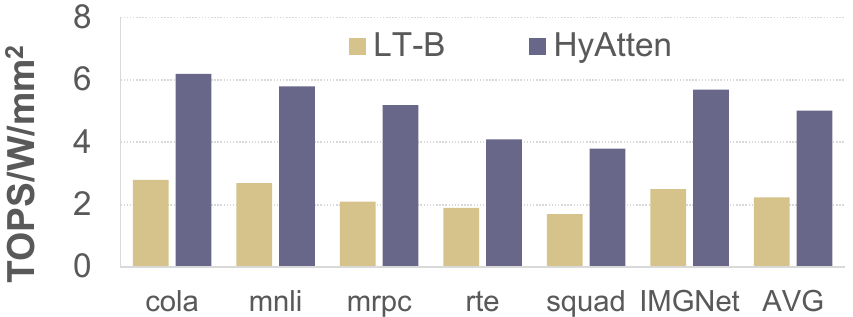}
\end{minipage}
}%
\centering
\caption{(a) Performance per unit area, and (b) energy efficiency per unit area}
\label{com_photonic}
\end{figure}

\section{Experimental Evaluation}

\subsection{Experimental Setup}
{\bf System Setup.} We modified an existing Python-based simulator~\cite{Zhu24} to evaluate the latency, power, area, and energy efficiency of HyAtten during Transformer inference. The area, leakage power, and access energy of the memory system are modeled using PCACTI~\cite{Shafaei14} in 14 nm. We model HBM with a bandwidth 1TB/s to supply data to the photonic system. The area and energy consumption of the digital die, including the softmax unit, MAU, and accumulator, are derived from SPICE circuit simulations~\cite{Corti20}. Similar to~\cite{Huang22, Huang23}, we scale the power of the ADC~\cite{Liu22} and PDAC~\cite{Sridarshini20} based on the bit-width and frequency requirements of the photonic PE. To further optimize, we replace one 8-bit ADC with 16 4-bit ADCs, as high-resolution ADCs can be constructed from multiple low-resolution units~\cite{Murmann2023}. Unlike conventional accelerators that share a single ADC across multiple arrays, HyAtten equips each array with 32 ADCs to minimize signal conversion latency. Table~\ref{tab:breakdown} provides a detailed breakdown of the device parameters used.

{\bf Models, datasets, training, and inference settings.} We evaluate the efficiency and accuracy of HyAtten using two widely recognized Transformer models: DeiT-T for vision tasks~\cite{Touvron21} and BERT-base for NLP tasks~\cite{Devlin18}. The models are tested on: ImageNet~\cite{Imagenet15} for vision and GLUE~\cite{Wang18glue} for NLP. Both weight and activation quantization are applied using low-bit precision~\cite{Esser19}. Additionally, noise-aware training is employed, with both encoding and systematic noise injected during training to reflect real-world conditions~\cite{Zhu24}.

\subsection{Compare to State-of-the-art Photonic Accelerator}
{\bf Baseline:} We select the base version of the Lightening-Transformer~\cite{Zhu24}, which features a 4-bit DPTC core, as the baseline system, denoted as LT-B. LT-B is a photonic-based Transformer accelerator that relies on high-resolution signal converters (e.g., 4-bit DAC and 8-bit ADC) to process all analog signals. To mitigate the area overhead associated with signal converters, LT-B configures each DPTC array with a single ADC. For a fair comparison, we use identical photonic device parameters for both LT-B and HyAtten. To eliminate the influence of chip area in our evaluation, we report performance and energy consumption normalized to per unit area.

{\bf Performance per unit area:} Figure~\ref{com_photonic} (a) dispalys the speedups achieved by HyAtten relative to the photonic baseline. Across all six datasets, HyAtten delivers a 9.8$\times$ speedups per unit area. This improvement can be attributed to two key factors. First, HyAtten replaces each high-resolution ADC with multiple low-resolution ADCs, significantly reducing signal conversion latency without increasing chip area. Second, HyAtten employs a digital die to handle the 15\% of signals that require high resolution ADCs, thereby avoiding excessive signal conversion overhead while incurring only a small area penalty. In contrast, the baseline LT-B system relies on high-resolution ADCs for all signals, which inefficiently allocates resources to process the 85\% of signals that are low-resolution.

{\bf Energy efficiency per unit area:} Figure~\ref{com_photonic} (b) illustrates the energy savings achieved by HyAtten compared to the photonic baseline. When processing all datasets using the DeiT-T and BERT models, HyAtten demonstrates a 2.2$\times$ energy reduction (normalized to the same chip area). These energy savings primarily stem from HyAtten's ability to eliminate the significant energy overhead associated with high-resolution signal converters. While the baseline LT-B system uses high-resolution ADCs to convert all signals, leading to substantial energy consumption, HyAtten replaces these high-resolution ADCs with low-resolution ADCs and digital dies, which perform GEMM operations with considerably lower energy demands.

\begin{figure}[t]
\centering
\subfloat[]{
\begin{minipage}[t]{0.495\linewidth}
\centering
\includegraphics[width=1.7in]{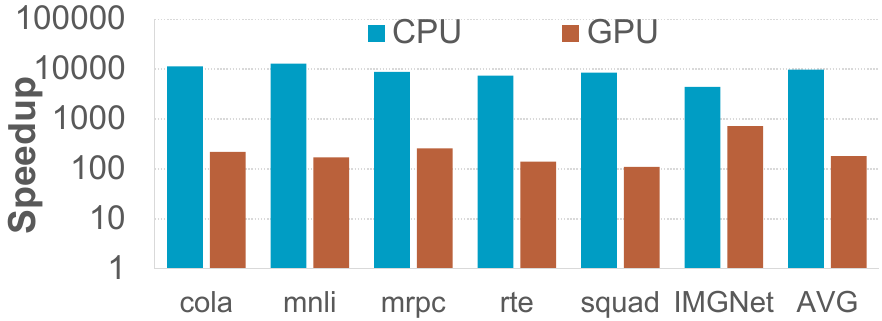}
\end{minipage}
}%
\subfloat[]{
\begin{minipage}[t]{0.495\linewidth}
\centering
\includegraphics[width=1.7in]{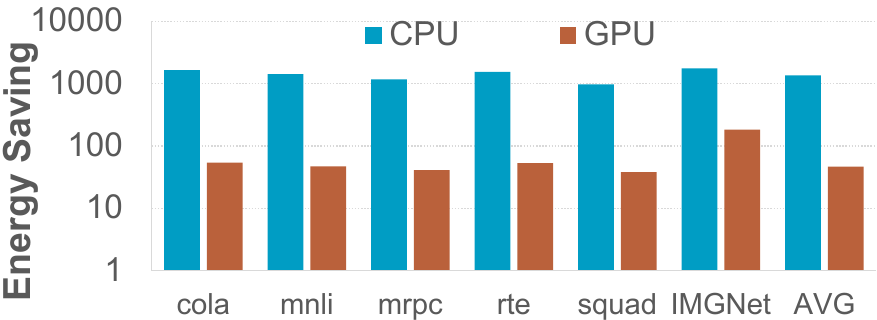}
\end{minipage}
}%
\centering

\caption{(a) HyAtten speedups compared to CPU and GPU, and (b) HyAtten energy saving compared to CPU and GPU}

\label{com_digital}

\end{figure}

\begin{table}[t]
\caption{{Accuracy Comparison}}

\centering
\label{tab:accuracy}
\setlength{\tabcolsep}{4.5mm}{
\begin{tabular}{cccc}
\hline
Model&\multicolumn{2}{c}{BERT}&DeiT \\
\hline
Datasets&GLUE (MRPC)&SQuAD&ImageNet-1K \\
\hline
Original&86.11&81.44&71.43 \\
LT-B&85.77&81.16&71.19 \\
HyAtten&85.89&81.22&71.27 \\
\hline
\end{tabular}}

\end{table}

\subsection{Compare to State-of-the-art Digital Accelerators}

In Figure~\ref{com_digital}, we compare HyAtten against, a single Nvidia A100 GPU and an Intel Core i7-9750H CPU, to highlight its significant performance and energy efficiency improvements. Figure~\ref{com_digital} (a) demonstrates that HyAtten delivers the highest performance, surpassing both CPU and GPU platforms. It achieves over 100$\times$ speedup per unit area relative to the A100 GPU, largely due to the high processing speed enabled by photonic computing. Figure~\ref{com_digital} (b) illustrates that HyAtten also exhibits superior energy efficiency, achieving over 50$\times$ greater efficiency per unit area compared to the GPU. This improvement is primarily attributed to the PDAC multiplexing within the DPTC array.

\subsection{Accuracy Comparison}
Table~\ref{tab:accuracy} reports the accuracy comparison between GPU, LT-B, and HyAtten, all running the same model, datasets, and bit-widths. HyAtten maintains an accuracy loss of less than 0.3\% compared to Transformers with the same bit-widths running on the GPU. Additionally, HyAtten demonstrates a 0.2\% accuracy improvement over LT-B. This accuracy improvement is primarily due to the integration of the digital die. As discussed in Section~\ref{sec:motiv}, high-resolution signals play a crucial role in maintaining model accuracy. The noise introduced by high-resolution ADCs in LT-B negatively affects these signals, leading to accuracy degradation. In contrast, HyAtten processes high-resolution signals using the noise-free digital die, thereby preserving model accuracy.

\begin{figure}[t]
\centering
\subfloat[]{
\begin{minipage}[t]{0.489\linewidth}
\centering
\includegraphics[width=1.5in]{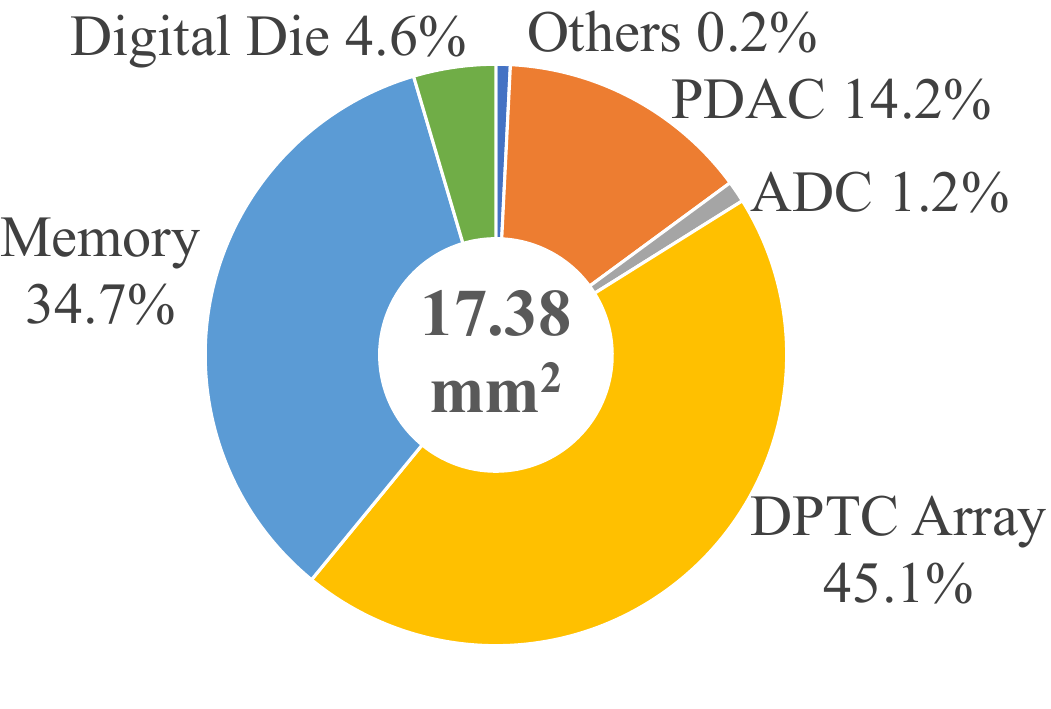}
\end{minipage}
}%
\subfloat[]{
\begin{minipage}[t]{0.489\linewidth}
\centering
\includegraphics[width=1.5in]{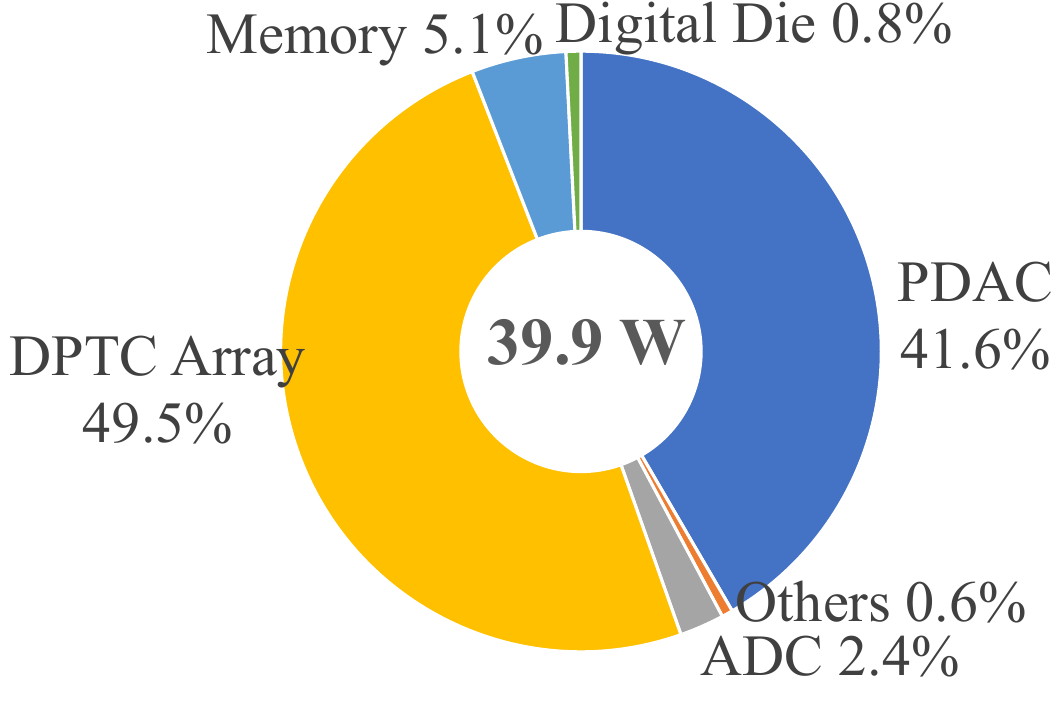}
\end{minipage}
}%
\centering

\caption{(a) Area breakdown, and (b) Power breakdown}

\label{breakdown}

\end{figure}
\subsection{System Efficiency Analysis}
{\bf Area Breakdown.} Figure~\ref{breakdown} (a) presents the area breakdown of HyAtten, which occupies a total area of 17.38mm$^2$. The DPTC array (including the {\em Mach-Zehnder modulator} (MZM), phase shifter, and photonic detector) accounts for the largest share at 45.1\%, followed by the memory system at 34.7\%, and the PDAC at 14.2\%. The remaining components, such as the ADC and digital die, contribute less than 10\% of the total area. Notably, HyAtten is designed with multiple ADCs for each DPTC array without increasing the overall area overhead. Additionally, the inclusion of a digital die to process around 15\% of the data introduces only a minimal increase in area.

{\bf Power Breakdown.} Figure~\ref{breakdown} (b) presents the power distribution for HyAtten, which consumes a total of 39.9W. The DPTC array and PDAC dominate the power usage, contributing about 49.5\% and 41.6\%, respectively. The remaining components, such as the memory system, ADC, and digital die, collectively account for less than 10\% of the total power consumption. Importantly, despite configuring additional ADCs for each DPTC array, the power overhead remains minimal. Additionally, the digital die, which processes around 15\% of the data, introduces only a negligible increase in overall power consumption.

\begin{figure}[t]
\centering
\subfloat[]{
\begin{minipage}[t]{0.495\linewidth}
\centering
\includegraphics[width=1.6in]{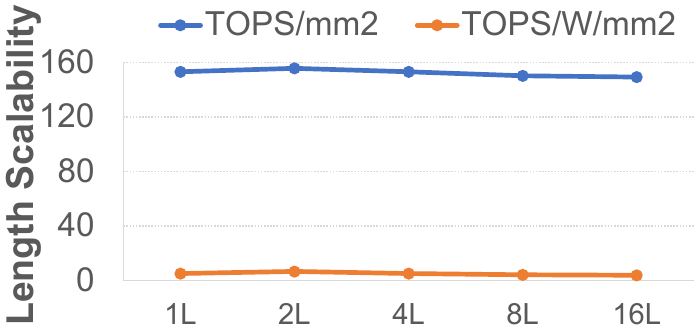}
\end{minipage}
}%
\subfloat[]{
\begin{minipage}[t]{0.495\linewidth}
\centering
\includegraphics[width=1.6in]{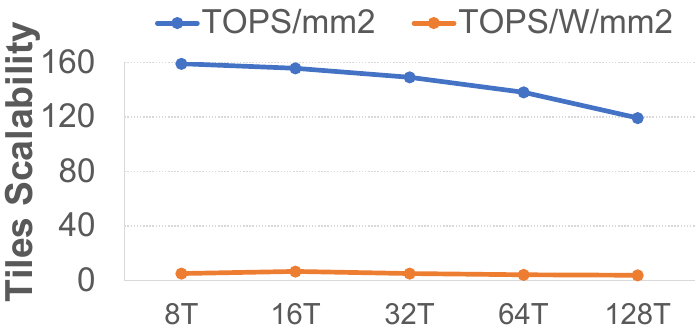}
\end{minipage}
}%
\centering

\caption{(a) Sequence length scalability, and (b) Tiles scalability}

\label{scalability}

\end{figure}

{\bf Scalability Discuss.} Figure~\ref{scalability} (a) presents the performance of HyAtten across various sequence lengths. As sequence length increases, HyAtten will process more submatrices, but the overhead associated with handling individual matrices remains relatively stable. Figure~\ref{scalability} (b) illustrates the impact of configuring different numbers of Tiles in HyAtten. While each Tile maintains a high internal throughput, increasing the number of Tiles leads to a rise in data transfer between HyAtten and the HBM, resulting in a reduction in overall throughput.

\section{Related Work}
{\bf Current Accelerators for Transformer.} {\em Field Programmable Gate Arrays} (FPGA) and {\em Application-specific integrated circuit} (ASIC) architectures are commonly used to accelerate sparse matrix-vector multiplication, such as graph processing~\cite{wangqg01, wangqg02, wangqg03, wangqg04, wangqg05, metanmp23}. Recently, researchers aim at accelerating attention mechanism with FPGA and ASIC architectures~\cite{Bai24, Sanger2021, You23}. Many memory-centric architectures are also proposed to accelerate attention mechanism, such as {\em processing in memory} (PIM)~\cite{cpsaa24, sadimm24, Li24, Zhou22, regnn2022, ReHarvest2024, liuc2024}. While these digital accelerators effectively reduce inference latency, traditional electrical computing platforms face significant limitations as transistor-based chips approach the boundaries of Moore's Law.

{\bf Emerging Photonic Architectures.} Integrated photonic accelerators present a promising alternative for accelerating deep neural networks, offering ultra-high speeds, extensive parallelism, and low energy consumption. Various optical systems have been explored to accelerate {\em convolutional neural networks} (CNNs)\cite{Feldmann21, shastri2021, shen2017deep, sunny2021crosslight, tait2017neuromorphic} and Transformers\cite{Zhu24}.

\section{Conclusion}
This paper presents HyAtten, a novel attention mechanism accelerator that leverages hybrid photonic and digital computing. HyAtten incorporates a photonic die with low-resolution ADCs to efficiently process low-resolution signals, while a digital die handles high-resolution signals without the overhead of signal conversion. Experimental results demonstrate that HyAtten achieves superior performance and energy efficiency with minimal accuracy loss.

\section{Acknowledgments}
This research is partially supported by the National Research Foundation, Singapore under its Competitive Research Program Award NRF-CRP23-2019-0003.

\bibliographystyle{IEEEtranS}
\bibliography{refs}

\end{document}